\documentclass{elsarticle}
\usepackage{times}
\usepackage{helvet}
\usepackage{courier}
\usepackage{amsfonts}
\usepackage{amsthm}
\usepackage{amsmath}
\usepackage{float}
\usepackage{algorithmic}
\usepackage{algorithm}

\begin{document}
\theoremstyle{plain}

\newcommand{\bit}{\{ 0, 1 \}}
\newcommand{\fullassign}{\bit^n}
\newcommand{\partassign}{\{0,1,*\}^n}
\newcommand{\Ints}{\mathbb{Z}}
\newcommand{\posInts}{\Ints^{+}}
\newcommand{\nonnegInts}{\Ints_{\geq 0}}
\newcommand{\boolfunc}[1][f]{\begingroup #1 \colon \bit^{n}\to \bit \endgroup}
\newcommand{\classfunc}[1][f]{\begingroup #1 \colon \bit^{n}\to \{1,\ldots,B\} \endgroup}
\newcommand{\utilfunc}[1][g]{\begingroup #1 \colon \partassign\to \nonnegInts \endgroup}
\newcommand{\OPT}{\mathsf{OPT}}
\newcommand{\ALG}{\mathsf{ALG}}

\newtheorem{claim}{Claim}
\newtheorem{theorem}{Theorem}
\newtheorem{lemma}{Lemma}
\newtheorem{corollary}{Corollary}
\newtheorem{proposition}{Proposition}

\title{The Stochastic Boolean Function Evaluation Problem for Symmetric Boolean Functions}
\author[dg]{Dimitrios Gkenosis}
\address[dg]{Department of Informatics and Telecommunications, University of Athens, Athens, Greece}
\author[ng]{Nathaniel Grammel}
\address[ng]{Department of Computer Science, University of Maryland, College Park, Maryland, USA}
\author[lh]{Lisa Hellerstein\corref{cor1}}
\cortext[cor1]{Corresponding author}
\ead{lisa.hellerstein@nyu.edu}
\address[lh]{Department of Computer Science and Engineering,
NYU School of Engineering,
370 Jay St.,
Brooklyn, NY 11201, USA}
\author[dk]{Devorah Kletenik}
\address[dk]{Department of Computer and Information Science,
Brooklyn College,
City University of New York,
2900 Bedford Avenue,
Brooklyn, NY 11210, USA
}

\begin{frontmatter}
\begin{abstract}
We give two approximation algorithms solving the Stochastic Boolean Function Evaluation (SBFE) problem for symmetric Boolean functions\footnote{\emph{Arxiv version note: Preliminary versions of these results appeared on Arxiv in arXiv:1806.10660. That paper contains results for both arbitary costs and unit costs. This paper considers only arbitrary costs.}}.  The first is an $O(\log n)$-approximation algorithm, based on the submodular goal-value approach
of Deshpande, Hellerstein and Kletenik. 
Our second algorithm, which is simple, is based on the algorithm solving the SBFE problem for $k$-of-$n$ functions, due to Salloum, Breuer, and Ben-Dov.  It achieves a 
$(B-1)$ approximation factor, where $B$ is the number of blocks of 0's and 1's in the standard vector representation of the symmetric Boolean function. 
As part of the design of the first algorithm, we prove that the goal value of any symmetric Boolean function is less than $n(n+1)/2$. 
Finally, we give an example showing that for symmetric Boolean functions, minimum expected verification cost and minimum expected evaluation cost are not necessarily equal.  This contrasts with a previous result, given by Das, Jafarpour, Orlitsky, Pan and Suresh, which showed that equality holds in the unit-cost case.
\end{abstract}
\end{frontmatter}


\section{Introduction}
\label{sec:intro}
In the Stochastic Boolean Function Evaluation (SBFE) problem, we are given (the representation of) a Boolean function $f(x_1, \ldots, x_n)$ that must be evaluated on an initially unknown assignment to the variables $x_i$.  The value of $x_i$ in this assignment can only be obtained by performing a test, which has an associated cost $c_i > 0$.  The probability that $x_i = 1$ is $p_i$, where $0 < p_i < 1$ and the $n$ tests are independent. Testing must continue until the outcomes of the performed tests are sufficient to determine the value of $f$. The problem is to determine the (adaptive) order in which to perform the tests, so as to minimize expected testing cost. (See Section~\ref{sec:defs} for a formal definition.)

There is an elegant polynomial-time algorithm that solves the SBFE problem when $f$ is a {\em $k$-of-$n$ function}, that is, a Boolean function whose output is 1 iff at least $k$ of its inputs $x_i$ are equal to 1. The original version of the algorithm, with its analysis, is due to Salloum, Breuer, and (independently) Ben-Dov
~\cite{Salloum79,BenDov81,SalloumBreuer84,Chang,Salloum97}.  

In this paper we consider the SBFE problem for a superclass of the $k$-of-$n$ functions, namely the class of {\em symmetric} Boolean functions.  A Boolean function is symmetric if its output depends only on the {\em number} of its inputs $x_i$ that are equal to 1.  
A symmetric Boolean function $f$ is represented by a vector of length $n+1$ indexed from 0 to $n$, which we call its {\em value vector}.  Position $j$ of the value vector contains the value of $f$ on input assignments $x$ containing exactly $j$ 1's. For example, if $f$ is the majority function on $n=3$ variables, then its value vector, indexed from 0 to 3, is $[0,0,1,1]$.

We note that Cicalese et al.\ previously 
presented a simple algorithm for symmetric Boolean function evaluation in a deterministic, on-line setting, where the goal is to minimize a worst-case
competitive ratio~\cite{cicalese2011competitive}. Their results do not apply to the SBFE problem.

{\bf Approximation algorithms and open questions:}
We present two approximation algorithms solving the SBFE problem for symmetric Boolean functions.  The first uses the {\em goal value approach} of Deshpande et al.~\cite{DeshpandeGoalValue} (Section~\ref{sec:goalvalue}).
It achieves an $O(\log n)$-approximation.
The second is a simple algorithm whose approximation factor is $(B-1)$, where $B$ is the number of ``blocks'' of consecutive 0's and consecutive 1's in the value vector for $f$.  For example, the value vector $[0,0,1,1,0]$ has two blocks of 0's and one block of 1's,  so $B=3$.
The $(B-1)$-approximation algorithm uses the $k$-of-$n$ evaluation algorithm of Salloum, Breuer, and Ben-Dov as a subroutine.  
Which approximation factor is smaller, $(B-1)$ or $O(\log n)$, depends on the relationship between $B$ and $n$ for the function in question. 

To achieve the approximation bound for the first algorithm, we prove a new structural result on symmetric Boolean functions: we show that the (submodular) {\em goal value} of a  symmetric Boolean function is upper bounded by $n(n+1)/2$.
This bound is almost tight, because for even $n$, the
goal value of the $k$-of-$n$ function for $k=n/2$ is exactly $(n/2)(n/2+1)$~\cite{bachetalgoalvalue}.


It remains an open question whether the SBFE problem for symmetric Boolean functions is NP-hard.
It is also open whether there is a polynomial-time constant-factor approximation algorithm.  

We note that the SBFE problem is known to be \textsf{NP}-hard for certain classes of Boolean formulas, including 
linear threshold formulas~\cite{heuristicLeastCostCox} and monotone DNF (or CNF) formulas~\cite{allen2017evaluation}.
The SBFE problem for linear threshold formulas has a polynomial-time constant-factor approximation algorithm, but if $P \neq NP$, the SBFE problem for monotone DNF (CNF) formulas has no such approximation algorithm~\cite{DeshpandeGoalValue,allen2017evaluation}.

Any symmetric Boolean function with $B=2$ must be either a $k$-of-$n$ function or the negation of a $k$-of-$n$ function.  Hence the SBFE problem for symmetric Boolean functions with $B=2$ can be solved exactly in polynomial time.  As we discuss below, there are polynomial-time exact algorithms solving the SBFE problem for 
some specific symmetric Boolean functions with $B=3$.
However, even in the unit-cost case, it is an open question whether there is a polynomial-time algorithm solving the SBFE problem for arbitrary symmetric Boolean functions with $B=3$. 

{\bf Evaluation vs. verification:}
The correctness of the algorithm solving the SBFE problem for $k$-of-$n$ functions, due to Salloum, Breuer,  and Ben-Dov, is based on a relationship between the evaluation problem and a related  verification problem.  
Intuitively, in the verification problem, you are given the same inputs as in the evaluation problem, and you are also given the value of $f(x)$.   You need only perform enough tests to verify that the given $f(x)$ value is correct.  
The correctness of the Salloum-Breuer-Ben-Dov algorithm for $k$-of-$n$ functions is based on the fact that for
$k$-of-$n$ functions, 
 optimal expected evaluation cost is equal to optimal expected verification cost 
 (cf.~\cite{unluyurtBorosDoubleRegular}).
 %

Subsequently, Das et al.\ showed that, in the special case of unit-costs (i.e., $c_i=1$ for all $i$),
equality of optimal evaluation and verification costs holds for {\em all} symmetric Boolean functions~\cite{Dasetal12}.
(Their work was inspired by work of Kowchik and Kumar, who re-discovered the unit-cost version of the Salloum-Breuer-Ben-Dov algorithm~\cite{kowshikkumar13}.)

The work of Das et al.\ did not address the question of whether  equality of optimal expected evaluation and verification costs holds for all symmetric Boolean functions when costs are arbitrary.  
We give a counterexample showing that it does not hold.

\bigskip
Preliminary versions of the results in this paper appeared previously in a conference paper~\cite{gkenosisetal18}.  
That paper also contained results for the unit-cost version of the SBFE problem, including a polynomial-time 4-approximation algorithm solving the SBFE problem for arbitrary symmetric Boolean functions in the unit-cost case.  Subsequent analysis has shown that the algorithm actually achieves a 2-approximation ~\cite{GHKL20}.

\section{Preliminaries}
\label{sec:defs}

An \emph{adaptive evaluation strategy} for a Boolean function $f(x_1, \ldots, x_n)$ is a sequential order in which to ``test'' the variables $x_i$ of $f$, so as to determine the value of $f$ on an initially unknown assignment $x \in \{0,1\}^n$.  Testing $x_i$ reveals its value. 
The choice of the next test can  depend on the outcomes of the previous tests.  

An adaptive evaluation strategy for $f$ corresponds to a {\em Boolean decision tree computing $f$}.  Each internal node of such a tree is labeled with a variable $x_i$ of $f$, and has two children, one corresponding to $x_i=1$ and the other to $x_i=0$.  Each leaf of the tree is labeled either 0 or 1.  An assignment $x \in \{0,1\}^n$ induces a root-leaf path in the tree, determined by the values of the $x_i$  in $x$.  The leaf at the end of that path is labeled with the value of $f(x)$.   

We do not require an SBFE algorithm to output the entire decision tree corresponding to the computed adaptive evaluation strategy, as that tree could be of exponential size.   It is sufficient for the algorithm to implement the strategy by sequentially computing the next test to perform, in an on-line fashion, and finally outputting the value of $f(x)$. 

Consider fixed values for the costs $c_i$ (where $c_i > 0$) and probabilities $p_i$ (where $0 < p_i < 1$) for the $n$ variables $x_i$.  We formally define the expected costs of adaptive evaluation and
verification strategies for $\boolfunc$ as follows.  
Given an adaptive evaluation strategy $\mathcal{A}$ for
$f$, and an assignment $x \in \fullassign$, we use
$C(\mathcal{A},x)$ to denote the sum of the costs $c_i$ of the tests performed in using $\mathcal{A}$ on $x$.  The expected cost of $\mathcal{A}$ is $\sum_{x \in \fullassign} C(\mathcal{A},x)p(x)$,
where $p(x) =\prod_{i=1}^n p_i^{x_i}(1-p_i)^{1-x_i}$ is the probability of $x$.  We say that $\mathcal{A}$ is an {\em optimal} adaptive
evaluation strategy for $f$ if it has minimum possible expected cost.

A {\em partial assignment} is a vector $b\in\partassign$. 
For $\ell \in \{0,1\}$, we use
$N_{\ell}(b)$ to denote $|\{i \mid b_i=\ell\}|$, the number of entries of $b$ that are set to $\ell$.
A partial
assignment represents the information that is known while performing
tests. Specifically, for a partial assignment $b\in\partassign$, $b_{i}=*$ indicates that the value of $x_i$ is still unknown, 
otherwise $b_i$ equals the
outcome of the test on $x_i$. 

We use $f^b$ to denote the restriction of function $f(x_1, \ldots, x_n)$ to the bits $i$ with $b_i = *$,
produced by fixing the remaining bits $i$ according to their values $b_i$.
We call $f^b$ the function {\em induced from $f$ by partial assignment $b$}.

An assignment $b'\in\{0,1\}^n$ is an \emph{extension} of a partial assignment $b \in \partassign$,
written $b'\succeq b$, if $b'_{i} = b_{i}$ for all $i$ such that
$b_{i} \neq *$.

A partial assignment $b \in \partassign$ is a {\em certificate} for a Boolean function $\boolfunc$ if $f(a)$ has the same value for all $a \in \fullassign$ such that $a \succeq b$.

For $\ell \in \{0,1\}$, let $\mathcal{X}_{\ell}=\{x\in\fullassign \mid f(x)=\ell\}$.
An {\em adaptive verification strategy} for $f$ consists of two adaptive evaluation strategies $\mathcal{A}_{\ell}$ for $f$, one for each ${\ell} \in \{0,1\}$.   The expected cost of the verification strategy is 
$\sum_{{\ell} \in \{0,1\}} \left ( \sum_{x \in \mathcal{X}_{\ell}}C(\mathcal{A}_{\ell},x)p(x) \right )$
and it is optimal if it minimizes this expected cost.

If $\mathcal{A}$ is an evaluation strategy for $f$, we call $\sum_{x \in \mathcal{X}_{\ell}} C(\mathcal{A},x)p(x)$ the $\ell$-cost of $\mathcal{A}$.
For $\ell \in \{0,1\}$, we say that $\mathcal{A}$ is {\em $\ell$-optimal} if it has minimum possible $\ell$-cost.
In an optimal verification strategy for $f$, each component evaluation strategy
$\mathcal{A}_{\ell}$ must be $\ell$-optimal.

A Boolean function $\boolfunc$ is \emph{symmetric} if its output on $x \in \{0,1\}^n$ depends only on $N_1(x)$, the number of 1's in $x$.
The \emph{value vector} for such a function $f$ 
is the $n+1$ dimensional vector $R$, indexed from $0$ to $n$,
whose $j$th entry $R[j]$ is the value of $f$ on inputs $x$ such that $N_1(x)=j$.
We partition the value vector $R$ into
\emph{blocks}. A
block is a maximal subvector of $R$ such that entries of the
subvector have the same value. Using $B$ to denote the number of blocks
of the value vector, we define $\alpha_1, \ldots \alpha_B$ to be the
minimum indices of each of the blocks, where
$0 = \alpha_1 < \alpha_2 < \ldots < \alpha_B$, and we define
$\alpha_{B+1} = n+1$.  
Block $i$ is the subvector of $R$ containing the entries indexed by the elements in $[\alpha_i, \alpha_{i+1})$.

We say that
an assignment $x$ {\em belongs to block $j$} if $\alpha_j \leq N_1(x) < \alpha_{j+1}$.  If $x$ belongs to block $j$, then $f(x)$ is equal to
$R[\alpha_j] = R[\alpha_j+1] = \ldots = R[\alpha_{j+1}-1]$.


A function $\utilfunc$ is {\em monotone} if $g(b') \geq g(b)$ whenever $b' \succeq b$.  It is {\em submodular} if for $b' \succeq b$, $i$ such that $b'_i = b_i = *$, and $\ell \in \{0,1\}$, we have $g(b'_{i \leftarrow \ell}) - g(b') \leq g(b_{i \leftarrow \ell}) - g(b)$.  Here $b_{i \leftarrow \ell}$ denotes the partial assignment produced from $b$ by setting $b_i$ to $\ell$, and similarly for $b'_{i \leftarrow \ell}$.

\section{Exact algorithms for special classes of symmetric functions}
\label{sec:exact}
Before  presenting our approximation algorithms, we describe exact algorithms solving the SBFE problem for some special classes of symmetric functions.

It is well-known that if $f$ is the Boolean OR function, then it is optimal to test the variables $x_i$ in nondecreasing order of the ratio $c_i/p_i$
(cf.~\cite{unluyurtReview}). Dually, if $f$ is Boolean AND, it is optimal to test in nondecreasing $c_i/(1-p_i)$ order. 

The Salloum-Breuer-Ben-Dov algorithm solving the SBFE problem for $k$-of-$n$ functions is recursive and works as follows.  Suppose $0 < k \leq n$.
Create two permutations of the $x_i$'s, one in nondecreasing order of the ratio $c_i/p_i$, and one in nondecreasing order of the ratio $c_i/(1-p_i)$.  By the pigeonhole principle, there must exist a variable $x_i$ that appears within the first $k$ variables of the first permutation, and within the first $n-k+1$ variables of the second permutation.  Find such a variable $x_i$ and test it.  If $x_i=1$, this reduces the problem to a $(k-1)$-of-$(n-1)$ evaluation problem, and if $x_i=0$, it reduces the problem to a $k$-of-$(n-1)$ evaluation problem.  Solve the reduced  problem recursively.  Assuming $0 \leq k \leq n$ at the start,
the base cases are
where $k=0$, implying that the value of $f$ must be 1, and where $k > n$, implying that the value of $f$ must be 0.

The correctness of this algorithm relies on the relationship between the verification and evaluation problems for $k$-of-$n$ functions (cf.~\cite{unluyurtBorosDoubleRegular}), as discussed in Section~\ref{append:verification}.

We note here that an almost identical algorithm solves the SBFE problem for 
{\em exactly-$k$ functions}, which output 1 iff
exactly $k$ of their inputs are 1. 
The only real difference in the algorithm 
is that instead of a base case for $k=0$, there is a base case for $k=-1$, implying that the value of $f$ is 0.  
The correctness proof is nearly identical to that for the $k$-of-$n$ algorithm.
The unit-cost version of the algorithm for exactly-$k$ functions was previously introduced by Acharya al.~\cite{Acharyaetal11}.
(They used the name
``delta functions'' to refer to the exactly-$k$ functions. A long, but more descriptive, name for them would be ``exactly-$k$-of-$n$ functions.'')

The value vector for an exactly-$k$ function contains exactly one 1, so $B = 3$ 
for any exactly-$k$ function where $1 \leq k \leq n-1$.  

Another interesting example of a symmetric function with $B=3$ is the {\em consensus} function.  The output of the  consensus function is 1 iff all of its inputs are equal, so its value vector has 1's only in its first and last positions.    
There is a polynomial-time exact algorithm that solves the SBFE problem for the consensus function, and for its complement, the not-all-equal function. 
We presented the unit-cost version of the algorithm in~\cite{GHKL20};
the extension to arbitrary costs (which we presented in our conference paper~\cite{gkenosisetal18}) is straightforward.

\section{Goal value and Adaptive Greedy}
~\label{sec:goalvalue}
The first algorithm we present for evaluating arbitrary symmetric functions
uses the goal value approach of Deshpande et al.~\cite{DeshpandeGoalValue}. (They called it the $Q$-value approach.)
The idea behind the approach is to solve the SBFE problem by reducing it to a (binary-state) Stochastic Submodular Cover problem. 
The latter problem is similar to the SBFE problem, with the following differences.  Instead of being given a Boolean function $f$ to evaluate, you are given (an oracle for) a monotone, submodular utility function $g(x_1, \ldots, x_n)$, where $g:\partassign \rightarrow \mathbb{R}_{\geq 0}$.
For simplicity, we will assume in what follows that $g$ is integer-valued, so
in fact, $\utilfunc$.
The function 
$g$ has the property that for all $a \in \fullassign$, the value of $g(a)$ is  equal to some common value $Q \in \nonnegInts$.  We call $Q$ the goal value of $g$.
Instead of performing tests until the value of a Boolean function $f$ can be determined, tests must be performed until the partial assignment $b$ representing the test outcomes so far satisfies $g(b) = Q$.
The problem is then to compute an adaptive strategy that minimizes expected testing cost.

To reduce the SBFE problem to the Stochastic Submodular Cover problem, 
we take the Boolean  function $\boolfunc$ that is to be evaluated in the SBFE problem
and use it to construct a utility function $\utilfunc$.  The function $g$ must be a {\em goal function} for $f$, meaning that it satisfies the following properties:
\begin{itemize}
    \item $g$ is submodular
    \item $g$ is monotone
    \item $g(*,\ldots,*)=0$
    \item there  exists  a value $Q \in \nonnegInts$ such that for all $b \in \partassign$,  $g(b) = Q$ iff $b$ is a certificate for $f$. 
\end{itemize}
For probabilities $p_i$ and costs $c_i$, finding an adaptive strategy of minimum expected cost for achieving goal utility  $Q$ (as measured by $g$) is then equivalent to finding an optimal adaptive  evaluation strategy for $f$.

The Adaptive Greedy algorithm of Golovin and Krause~\cite{golovinKrause} is an approximation algorithm for the Stochastic Submodular Cover problem.  To choose the variable $x_i$ to test next, it uses the following greedy rule: Choose the $x_i$ whose test outcome maximizes the expected increase in utility, per unit cost (with respect to $g$, the $p_i$, and the $c_i$). 
There are a number of different proofs showing that
Adaptive Greedy achieves an $O(\log Q)$ approximation bound for the Stochastic Submodular Cover problem~\cite{hellersteinKletenikParthasarathy21,DeshpandeGoalValue,hellerstein2018revisiting,imetal}.\footnote{The tightest of these $O(\log Q)$ bounds is the $(\ln Q + 1)$ bound due to Hellerstein et al.~\cite{hellersteinKletenikParthasarathy21}.
 An earlier proof of a $(\ln Q+1)$ bound was found to have an error~\cite{golovinKrause,nanSaligrama,golovinKrauseArxivv5}.  
 A recent $O(\log Qn)$ bound, due to Esfandiari et al.~\cite{esfandiarietal19}, 
 also applies to a generalization of the Stochastic Submodular Cover problem.}
 
Thus once a submodular goal function $g$ is constructed for Boolean function $f$, running Adaptive Greedy on $g$ results in an $O(\log Q)$-approximation to the optimal adaptive strategy for evaluating $f$.  
The challenge in the goal value approach is to construct $g$ so that its goal value $Q$ is small, resulting in a small approximation factor. 
This is not possible for all 
classes of Boolean functions $f$.  
The {\em goal value of a Boolean  function $f$} is the minimum goal value of any submodular goal function $g$ for $f$. 
The goal value of every Boolean function is upper bounded by $2^n$~\cite{bachetalgoalvalue}.
While we show that symmetric functions have goal value polynomial in $n$, 
some classes of Boolean functions have goal value exponential in $n$~\cite{DeshpandeGoalValue,bachetalgoalvalue}.  

\section{An $O(\log n)$-approximation algorithm based on goal value}

We present the $O(\log n)$-approximation algorithm for the SBFE problem for symmetric functions, using the goal value approach.  To implement this approach, we construct 
a submodular goal function $g$ for $f$.  
The construction is in the proof of the following theorem, which gives an upper bound on the goal value of any symmetric function.

\begin{theorem}
\label{thm:goalval}
The goal value of any symmetric Boolean function $f(x_1, \ldots, x_n)$ is strictly less than $n(n+1)/2$.
\end{theorem}

\begin{proof}
Let  $f(x_1, \ldots, x_n)$ be a symmetric Boolean function. We construct a submodular goal function $g$ for $f$ using its value vector $R$.  The construction is based on a graph $G$, defined as follows.
The graph has $n+1$ vertices, $v_0, \ldots, v_n$, where $v_i$ corresponds to position $i$ of $R$.
Let $B$ be the number of blocks in $R$.
Partition the vertices into $B$ subsets, where each subset contains the
vertices corresponding to the positions contained in a single block.
The graph $G$ is the complete $B$-partite graph induced by this partition,
so there is an edge from $v_i$ to $v_j$ iff positions $i$ and $j$ are in different blocks of $R$.

We construct a goal function $g$ for $f$ that assigns a value to each partial assignment $b \in \partassign$.  
For $b \in \partassign$, let $V(b)$ be the set of vertices $v_i$ of $G$ such that either $i < N_1(b)$ or $i > n-N_0(b)$.
Say that $b$ {\em covers} an edge of $G$ if $V(b)$ contains at least one of its endpoints.  
Let $S(b)$ be the set of edges of $G$ that are covered by $b$.
We define $g(b) = |S(b)|$.

We now argue that $g$ is a goal function for $f$.
Clearly $g(*,\ldots,*) = 0$.  If $b' \succeq b$ then $N_1(b') \geq N_1(b)$ and $N_0(b') \geq N_0(b)$.  Thus
$S(b) \subseteq S(b')$, so $g$ is monotone.

Every partial assignment $b \in \partassign$ induces a function $f^b$ of $f$ which is a symmetric function on the variables $x_i$ for which $b_i=*$.  The value vector of $f^b$ is produced from $R$ by deleting its first $N_1(b)$ entries and its last $N_0(b)$ entries (which are disjoint, since $N_0(b) + N_1(b) \leq n$).  The function $f^b$ is constant iff all entries in its value vector are equal.  That is, $b$ is a certificate for $f$ iff removing the first $N_1(b)$ and last $N_0(b)$ entries from $R$ results in the remaining entries all being members of a single block of $R$.  This latter property holds iff $S(b)$ is the set of all edges of $G$.  
Thus $g(b)$ is equal to the total number of edges of $G$ iff $b$ is a certificate of $f$.

Finally, we show that $g$ is submodular.  Consider
some $b \in \partassign$ and $i \in \{1, \ldots, n\}$
such that $b_i = *$. 
Let $\ell=1$; an analogous argument holds for $\ell=0$.
Consider
$g(b_{i \leftarrow \ell}) - g(b)$, which is equal to
$|S(b_{i \leftarrow \ell}) \backslash S(b)|$.
Let $s = N_1(b)$.  Vertex $v_s$
is the only vertex in $V(b_{i \leftarrow \ell})$ that is not in $V(b)$.
It follows that the edges in
$S(b_{i \leftarrow \ell}) \backslash S(b)$
are precisely the pairs $\{v_s,v_t\}$ where
$t$ is in the set $I(b) := \{N_1(b)+1, \ldots, n-N_0(b)\}$, and $s$ and $t$ are in different blocks of $R$.  

Now consider $b' \in \partassign$ such that $b' \succeq b$ and $b'_i=*$.
Let $s' = N_1(b')$.

Consider an edge $\{v_{s'},v_t\}$ in 
$S(b'_{i \leftarrow \ell}) \backslash S(b')$.  Clearly $t \in I(b')$.  
Since $N_1(b) \leq N_1(b')$ and $N_0(b) \leq N_0(b')$, we have 
$t \in I(b)$ as well.
Since $s'$ and $t$ are in different blocks of $R$, and $s \leq s'$, it follows that $s$ and $t$ are also in different blocks of $R$.  Thus for each edge $\{v_{s'},v_t\}$ in 
$S(b'_{i \leftarrow \ell}) \backslash S(b')$ there is a corresponding edge $\{v_s,v_t\}$ in
$S(b_{i \leftarrow \ell}) \backslash S(b)$.  It follows that 
$|S(b'_{i \leftarrow \ell}) \backslash S(b')| \leq 
|S(b_{i \leftarrow \ell}) \backslash S(b)|$ and therefore
$g(b'_{i \leftarrow \ell}) - g(b')
\leq g(b_{i \leftarrow \ell}) - g(b)$.
Since an analogous argument holds if $\ell=0$, $g$ is submodular.

The number of edges in $G$ is maximized when each position of $R$ is in its
own block.  In this case, the goal value of the constructed $g$ is $n(n+1)/2$.
This implies that the goal value of any symmetric function is at most $n(n+1)/2$.

To see that it is strictly less than $n(n+1)/2$, note that
there are only two symmetric functions $f$ for which each position of $R$ is in its own block: the parity function and its complement.  
For each of these functions, the above construction does not achieve minimum possible goal value.  The goal value of these functions is $n$, and it is achieved by the utility function $g(b) = N_0(b) + N_1(b)$ ~\cite{bachetalgoalvalue}.   Thus every symmetric Boolean function has goal value strictly less than $n(n+1)/2$.
\end{proof}

The construction in the above proof generalizes a previous construction for $k$-of-$n$ functions.  In that case, the graph $G$ is bipartite and the constructed function achieves minimum possible goal value for the $k$-of-$n$ function~\cite{bachetalgoalvalue}. It is an open problem to give a construction that achieves minimum possible goal value for every symmetric function.

Having described the construction, we prove the following theorem.

\begin{theorem}
There is a polynomial-time $O(\log n)$-approximation algorithm for the 
SBFE  problem for symmetric Boolean functions.
\end{theorem}

\begin{proof}
The input to the SBFE problem for symmetric Boolean functions is the value vector $R$ for the symmetric function $f$ that is to be evaluated.

The algorithm uses $R$ to construct the graph $G$ defined in the proof of Theorem~\ref{thm:goalval}.  Once $G$ is constructed, the value of the associated utility function $g$ 
can be easily computed on any given partial assignment.  
The algorithm runs the Adaptive Greedy algorithm of Golovin and Krause on $g$ to determine the order in which to perform the tests.  Let $b \in \partassign$ be the partial assignment representing the outcomes of all the tests performed by Adaptive Greedy.  By the proof of Theorem~\ref{thm:goalval}, the entries $R[N_1(b)], \ldots, R[n-N_0(b)]$ are all equal to the desired value $f(x)$.
The algorithm outputs one of them, e.g., $R[N_1(b)]$.

Let $Q$ denote the goal value of $g$.
As shown in the proof of Theorem~\ref{thm:goalval}, it is at most $n(n+1)/2$.
Because Adaptive Greedy is an $O(\log Q)$-approximation algorithm for Stochastic Submodular Cover, the above algorithm achieves an approximation factor of $O(\log n)$.
\end{proof}

\section{The $(B-1)$-approximation algorithm}
\label{sec:B-1approx}

The $(B-1)$-approximation algorithm for the SBFE problem is simple.
It runs the algorithm for evaluating $k$-of-$n$ function, due to Salloum, Breuer, and Ben-Dov, once for each of the $B-1$ values $\alpha_2, \ldots, \alpha_B$ associated with the value vector of $f$.  The run for $\alpha_j$ sets $k=\alpha_j$, and determines
whether the initially unknown assignment 
$x$ satisfies $N_1(x) \geq \alpha_j$, i.e., whether $x$ belongs to a block numbered $j$ or higher.  
Once this is done for the $B-1$ values $\alpha_2, \ldots, \alpha_B$,
it is straightforward to determine which block of the value vector contains $x$, and hence to determine the value of $f(x)$.

We present the pseudocode for this algorithm as Algorithm~\ref{alg:repeatedkofn} and
show it achieves a $(B-1)$-approximation.
In the pseudocode, we denote as $f_{i}$ the $k$-of-$n$
function with $k=\alpha_i$. We note that in different iterations of the for loop, the strategy that is executed in the body may choose a test that was already performed in a previous iteration.
The test does not actually have to be repeated, as the outcome can be stored after the first time the test is performed, and accessed whenever the test is chosen again.

\begin{algorithm}[ht]
  \caption{$(B-1)$-approximation algorithm}
  \label{alg:repeatedkofn}
  \begin{algorithmic}
    \FOR{$i\gets 2$ {\bfseries to} $B$}
    \STATE Using the $k$-of-$n$ evaluation algorithm of Salloum, Breuer, and Ben-Dov, perform tests to find the value of $f_{i}(x)$
    \ENDFOR
    \STATE if $f_i(x)=0$ for all $i > 1$, set $i^* = 1$~~// ~recall that $\alpha_1 = 0$
    \STATE else set $i^{*}\gets \max\{i\mid f_{i}(x)=1\}$ 
     \RETURN $R[\alpha_{i^{*}}]$
  \end{algorithmic}
\end{algorithm}

The  correctness of the algorithm follows easily
from the facts that $f_i(x) = 1$ iff $N_1(x) \geq \alpha_i$, and that
$\alpha_{i^{*}}\leq N_{1}(x) < \alpha_{i^{*}+1}$, and so
$f(x) = R[\alpha_{i^{*}}]$.

We now examine the expected cost of the strategy computed in
Algorithm~\ref{alg:repeatedkofn}. Let $C(f_{i})$ denote the expected cost of
evaluating $f_{i}$ using the optimal $k$-of-$n$ strategy. Let $\OPT$ be the expected cost of the optimal adaptive strategy for $f$.

\begin{lemma}
  \label{lem:kofnltopt}
  $C(f_{i})\leq \OPT$ for $i \in \{2, \ldots, B\}$.
\end{lemma}

\begin{proof}
Let $T$ be the decision tree corresponding to an optimal adaptive strategy for evaluating $f$.
Consider using $T$ to evaluate $f$ on an initially unknown input $x$.
When a leaf of $T$ is reached, we have discovered the values of some of the
bits of $x$.  Let $d$ be the partial assignment representing that knowledge.
Recall that $f^{d}$ is the function induced from $f$ by $d$.
The value vector of $f^{d}$ is a subvector of the value vector $R$ of $f$.
More particularly, it is the subvector stretching from index $N_1(d)$ of $R$
to index $n - N_0(d)$.
Since $T$ is an evaluation strategy for $f$,
reaching a leaf of $T$ means that we have enough information to determine $f(x)$.
Thus all entries of the subvector must be equal, implying that it is contained within a single
block of $R$.  We call this the block associated with the leaf.

For each block $i \geq 2$, we can create
a new tree $T'_{i}$ from $T$ which evaluates the
  function $f_{i}$.  We do this by relabeling the leaves of $T$: if the leaf is associated with block $i'$, then we label the leaf with output value 1 if $i' \geq i$, and with 0 otherwise.   $T'_{i}$ is an adaptive strategy for evaluating $f_{i}$.

  The expected cost of evaluating $f_{i}$
  using $T'_{i}$ is equal
  to $\OPT$, since the structure of the tree is unchanged from $T$
  (we've only changed the labels on leaves). Since $T'_{i}$ cannot do better
  than the optimal $k$-of-$n$ strategy, $C(f_{i}) \leq \OPT$.
\end{proof}

Lemma~\ref{lem:kofnltopt} yields the following theorem:
\begin{theorem}
There is a polynomial-time $(B-1)$-approximation algorithm for the 
SBFE  problem for symmetric Boolean functions.
\end{theorem}
\begin{proof}
Let $C(f_i, x)$ denote the cost of evaluating $f_i(x)$ using the optimal $k$-of-$n$ evaluation algorithm.  Then
the expected cost of Algorithm~\ref{alg:repeatedkofn} is $\sum_{x\in \bit^n} \sum_{i=2}^{B} C(f_i, x)p(x)$. 
Reversing the order of summation,
this is equal to $\sum_{i=2}^{B} \sum_{x\in \bit^n} C(f_i, x)p(x) = \sum_{i=2}^{B} C(f_i)$. 
Thus by Lemma~\ref{lem:kofnltopt}, the total cost of Algorithm~\ref{alg:repeatedkofn} is at most
$\sum_{i=2}^{B} C(f_{i}) \leq \sum_{i=2}^{B}\OPT = (B-1)\OPT$.
  \end{proof}

  We note that we could easily modify this algorithm
  to use binary search (to find the index $i^*$ of the block containing $x$), rather than sequential search.
  However, while this seeemingly would lead to an approximation bound of $O(\log B)$, we do not know how to prove that the modified algorithm actually achieves such a bound.
  The difficulty in simply adapting the previous analysis is as follows.
  Consider the expression for the expected cost of the previous algorithm, $\sum_{x\in \bit^n} \sum_{i=2}^{B} C(f_i, x)p(x)$.
   If the algorithm is modified to use binary search rather than sequential search, then for each $x$,
  the inner sum, $\sum_{i=2}^{B} C(f_i, x)p(x)$,
   can be replaced by a sum of  $O(\log B)$ terms of the form $C(f_i,x)p(x)$. However, these terms would not be the same for all $x$, because they depend on the execution of the binary search associated with $x$. This prevents us from reversing the order of summation, as we did in the previous argument, which prevents us from completing the proof and attaining the desired $O(\log B)$ bound.
   
\section{Verification vs. Evaluation}
\label{append:verification}

Recall the definitions associated with verification strategies and their costs, 
from Section~\ref{sec:defs}. 
Let $\boolfunc$ be a Boolean function with associated costs $c_i$ and probabilities $p_i$.
Let $\mathcal{V}(f)$ and $\mathcal{E}(f)$ denote the minimum possible expected verification cost and minimum possible expected evaluation cost, respectively, of $f$.

The correctness of the algorithm of Salloum, Breuer, and Ben-Dov solving the SBFE problem for $k$-of-$n$ functions is based on the following lemma.     

\begin{lemma}
\label{optimalverify}
\cite{BenDov81} 
Consider an instance of the SBFE problem.
If $f$ is a $k$-of-$n$ function, then
the evaluation strategy that tests the bits in nondecreasing order of the ratio
$c_i/p_i$, until the value of $f$ can be determined, is 1-optimal.
\end{lemma}

A dual lemma states that testing in nondecreasing order of $c_i/(1-p_i)$ is 0-optimal when $f$ is a $k$-of-$n$ function.  

It is obvious that $\mathcal{V}(f) \leq \mathcal{E}(f).$  
Before discussing this relationship further, we describe the proof of
correctness for the Salloum-Breuer-Ben-Dov algorithm, presented in Section~\ref{sec:exact}, for $k$-of-$n$ functions.  Since any 1-optimal strategy must perform at least $k$ tests before terminating, the strategy that tests in nondecreasing $c_i/p_i$ order is still 1-optimal if you permute the first $k$ tests in  the ordering arbitrarily.  Similarly, the strategy that tests in nondecreasing $c_i/(1-p_i)$ order is still 0-optimal if you permute the first $n-k+1$ tests arbitrarily.  The algorithm first tests an $x_i$ that appears within the first $k$ tests of the $c_i/p_i$ ordering, and also within the first $n-k+1$ tests of the $c_i/(1-p_i)$  ordering.  
Thus testing such an $x_i$ first is  consistent with both a 1-optimal and a 0-optimal strategy. 
This also holds for the recursive calls of the algorithm.  Thus the strategy produced by the algorithm is both 0-optimal and 1-optimal, and because $\mathcal{V}(f) \leq \mathcal{E}(f)$, it is an optimal evaluation strategy.  

The  above correctness proof also implies that for $k$-of-$n$ functions $f$, $\mathcal{V}(f) = \mathcal{E}(f)$.   

Das et al.~\cite{Dasetal12} showed that {\em in the  unit-cost case}, $\mathcal{V}(f) = \mathcal{E}(f)$ holds when $f$ is an arbitrary symmetric Boolean function $f$. 
Their proof also relied on showing that there exists a variable $x_i$ such that testing $x_i$ first is consistent with both a 1-optimal and a 0-optimal strategy.  However, the proof involves additional insights and arguments that were not needed for $k$-of-$n$ functions, and it does not imply a polynomial-time procedure to find such an $x_i$.     

We show that the result of Das et al.\ does not hold for arbitrary costs.
We describe a symmetric Boolean function $f$, and associated $p_i$ and $c_i$, for which the minimum expected cost of evaluation exceeds the minimum expected cost of verification. The description of the function is in the proof of the following theorem.  

\begin{theorem}
\label{thm:verif}
There exists a symmetric Boolean function $\boolfunc$, costs $c_i$, and probabilities $p_i$, such that $\mathcal{V}(f) < \mathcal{E}(f)$.
\end{theorem}
\begin{proof}
We give a function $f$ on $n = 4$ bits with $B = 3$.  The value vector of $f$ is $[0,1,1,0,0]$.
The costs and probabilities for the bits are given in Table~\ref{tab:vars}. 

\renewcommand{\arraystretch}{1}
\begin{table}[ht!]
\centering
\begin{tabular}{l l l}
\hline
bit & $p_i$ & cost \\
$x_1$ & 0.1 & 5000 \\
$x_2$ &  0.3 & 6000 \\
$x_3$ & 0.9 & 3000 \\
$x_4$ & 0.8 & 5000 \\
\hline
\end{tabular}
\caption{Costs and probabilities of bits}
\label{tab:vars}
\end{table}


Consider the evaluation tree for ${f}$ given in Figure~\ref{fig:opteval}; we denote it by $T$.  We assume left edges correspond to $x_i=0$ and right edges to $x_i=1$. 
So, for example, if $x \in \bit^4$ is such that $x_1=x_2=0$ and $x_3=x_4 = 1$, then 
$p(x)=(1-p_1)*(1-p_2)*p_3*p_4=.4536$ and the
sum of the tests performed by $T$ on $x$ is
$C(T,x)=c_3+c_2+c_1=14000$.
The expected cost of $T$ is $\sum_{x \in \fullassign} C(T,x)p(x)=14,618$.

In fact, 
$T$ is optimal, meaning that it has minimum possible expected cost over all adaptive strategies for evaluating $f$. Thus
$\mathcal{E}({f}) = 14,618$.
The optimality of $T$ 
can be shown by computing the expected cost of only 12 candidate trees, as follows.
For any evaluation strategy, if the outcome of the first test is 1, then the induced problem is to evaluate the function with value vector $[1,1,0,0]$; this new function is the negation of a 2-of-3 function. If the outcome of the first test is 0 and the outcome of the second test is 1, then the induced problem is to evaluate the function with value vector $[1,1,0]$.  This function is the negation of a 2-of-2 function.  
If the outcome of the first test is 0 and the outcome of the second test is also 0, then the induced problem is to evaluate the function with value vector $[0,1,1]$.  This is a 1-of-2 function.
Since we know the optimal evaluation strategies for $k$-of-$n$ functions (and their negations), to determine the optimal evaluation tree for $f$, we only need to determine which variables appear in the root of the tree and in its left child.  
We do this by trying all 12 choices for these two variables, and computing the expected costs of the associated trees.  The results are shown in Table~\ref{tab:comb}; the optimal expected cost is bolded. 

\begin{figure}[h!]
\centering
\includegraphics[scale=.65]{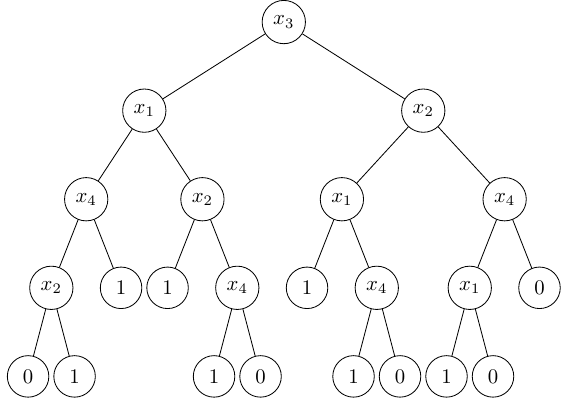}
\caption{Optimal evaluation tree for $f$}
\label{fig:opteval}
\end{figure}

\renewcommand{\arraystretch}{1}

\begin{table}[ht!]
\centering
\begin{tabular}{ccc}
\hline
root & left child & expected cost of tree \\

$x_1$ & $x_2$ & 15,529\\
$x_1$ & $x_3$ & 15,259 \\
$x_1$ & $x_4$ & 16,042 \\
$x_2$ & $x_1$ & 14,881 \\
$x_2$ & $x_3$ & 14,643 \\
$x_2$  & $x_4$& 15,616\\
$\mathbf{x_3}$ & $\mathbf{x_1}$ & \textbf{14,618} \\
$x_3$ & $x_2$ & 14,670\\
$x_3$ & $x_4$ & 14,623\\
$x_4$ & $x_1$ &  15,394\\
$x_4$ & $x_2$ & 15,616\\
$x_4$ & $x_3$ & 15,406\\
\hline
\end{tabular}
\caption{Possible evaluation trees for $f$ and their costs}
\label{tab:comb}
\end{table}

Now consider the problem of verifying $f$.
Recall that a verification strategy for $f$ consists of two evaluation strategies, one 
for assignments $x$ in $\mathcal{X}_{1}$
and one for assignments $x$ in
$\mathcal{X}_{0}$.
If $\mathcal{V}(f)$ were equal to $\mathcal{E}(f)$, then since $T$ is an optimal adaptive strategy for $f$, it would have to be $\ell$-optimal for each $\ell \in \{0,1\}$.  Otherwise, if $T$ were not $\ell$-optimal for some $\ell$, one could achieve an expected  verification cost lower than $\mathcal{E}(f)$ by
using an $\ell$-optimal tree for the assignments in $\mathcal{X}_{\ell}$ and the tree $T$ for the assignments in $
\mathcal{X}_{1-\ell}$.

In Figure~\ref{fig:optverif} we show a truncated version of an evaluation tree $T'$ for $f$ whose 1-cost is $\sum_{x \in \mathcal{X}_{\ell}} C(\mathcal{T'},x)p(x)=
\mathbf{10,241.8}$.
(In fact, this is the optimal 1-cost.)  
In the figure, X designates a leaf which is not reachable on any $x$ for which $f(x)=1$, and thus that node and its descendants in the original tree do not affect the 1-cost.

The 1-cost of the optimal evaluation tree $T$ in Figure~\ref{fig:opteval}
is $\mathbf{10,248.8}$.
Because the 1-cost of the tree
in Figure~\ref{fig:optverif} 
tree is less than the 1-cost of $T$,
$\mathcal{V}(f)< \mathcal{E}(f)$.

\begin{figure}[ht]
\centering
\includegraphics[scale=.65]{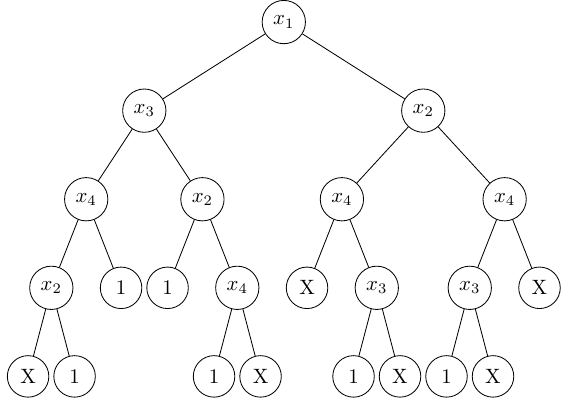}
\caption{Tree with optimal 1-cost}
\label{fig:optverif}
\end{figure}
\end{proof}

\section*{Acknowledgments}
Partial support for this work came from NSF Award IIS-1217968 (for all authors), from
NSF Award IIS-1909335 (for L. Hellerstein), from a PSC-CUNY Award, jointly funded by The Professional Staff Congress and The City University of New York (for D. Kletenik).
We thank Zach Pomerantz for experiments that gave us useful insights into the goal value of symmetric functions and the anonymous referees for their comments.

\bibliographystyle{elsarticle-harv}
\bibliography{miscrefs.bib,throughput.bib,soda5.bib}

\end{document}